\documentclass[useAMS,usenatbib]{mn2e}
\usepackage{times}
\usepackage{graphics,epsfig}
\usepackage{graphicx}
\usepackage{amssymb}

\title[Spherical accretion in a fractal medium]
{Critical properties of spherically symmetric accretion in a 
fractal medium}
\author[Roy and Ray]
{Nirupam Roy$^{1}$\thanks{nirupam@ncra.tifr.res.in}
and Arnab K. Ray$^{2}$\thanks{akr@iucaa.ernet.in}\\
$^{1}$National Centre for Radio Astrophysics, Tata Institute of
Fundamental Research, Post Bag 3, Ganeshkhind, Pune 411007, India\\
$^{2}$Inter--University Centre for Astronomy and Astrophysics, Post
Bag 4, Ganeshkhind, Pune 411007, India}

\begin{document}


\maketitle

\label{firstpage}

\begin{abstract}
Spherically symmetric transonic accretion of a fractal medium
has been studied in both the stationary and the dynamic regimes.
The stationary transonic solution is greatly sensitive to 
infinitesimal deviations in the outer boundary condition, but
the flow becomes transonic and stable, when its evolution 
is followed through time. The evolution towards transonicity 
is more pronounced for a fractal medium than what is it 
for a continuum. The dynamic approach also shows that there 
is a remarkable closeness between an equation of motion for 
a perturbation in the flow, and the metric of an analogue 
acoustic black hole. The stationary inflow solutions of a 
fractal medium are as much stable under the influence of 
linearised perturbations, as they are for the fluid continuum. 
\end{abstract}


\section{Introduction}
\label{sec1}

Accretion processes involve the non-self-gravitating flow dynamics
of astrophysical matter under the external gravitational influence
of a massive astrophysical object, like an ordinary star or a compact
object~\citep{fkr02}.
A paradigmatic model of an astrophysical accreting system is that
of spherically symmetric infall on to a central accretor. Ever since
the seminal work published by~\citet{bon52}, which effectively
launched the subject in the form in which it is recognised today,
the problem of spherically symmetric flows has been revisited time
and again from various angles~\citep{par58,par66,an67,bal72,mich72,
mes75,blum76,ms77,beg78,cos78,sb78,gar79,bri80,monc80,pso80,vit84,
bona87a,bona92b,td92,km94,mar95,tuf95,tmk96,zmt96,tmk97,ke98,das99,
malec99,ttsrcl99,das00,ds01,rb02,ray03,das04,rb05,gai06,mrd07,roy07}.
This abiding appeal of the spherically symmetric model is explained
by the fact that almost always it lends itself to an exact mathematical 
analysis, and in the process it allows a very clear insight to be had 
into the underlying physical principles. 

Ease of mathematical manipulations, however, is not the only reason why 
spherically symmetric flows are regularly invoked in accretion-related 
literature. The details of the physics of many astrophysical flows 
are, in fact, very faithfully described and understood with the help
of this relatively simple model. Accretion of the interstellar medium 
(ISM) is a case in point. 

While formal fluid dynamical equations in the Newtonian construct of 
space and time --- which would involve a momentum balance equation 
(with gravity as an external driving force), the continuity equation 
and an equation of state --- suffice to a great extent in 
shedding light on the accretion of the ISM, it must
at the same time be recognised that the ISM is not entirely to be 
seen as a fluid continuum. In fact, for many purposes essential to  
grasping the underlying details, the ISM is believed to possess a 
self-similar hierarchical structure over several orders of magnitude 
in scale~\citep{lars,falg92,heith}.
Direct H~{\sc i} absorption observations and interstellar scintillation
measurements suggest that the structure extends down to a scale of 
$10\, \mathrm{au}$~\citep{crov,lang,fais} and possibly even to  
sub-$\mathrm{au}$ scales~\citep{hill}.
Numerous theories have attempted to explain the origin, evolution 
and mass distribution of these clouds and it has been established, 
from both observations~\citep{elme} and numerical 
simulations~\citep{burk,kless,seme}, that the interstellar medium 
has a clumpy hierarchical self-similar structure with a fractal 
dimension in three-dimensional space. The main reason 
for this is still not properly understood, but it can be the consequence 
of an underlying fractal geometry that may arise due to turbulent 
processes in the medium.

A theoretical study of these astrophysical systems --- either a 
fluid continuum or a fractal structure --- will necessitate the 
application of the mathematical principles of nonlinear dynamics. 
This is the principal objective of this work. The physical processes 
in a fractal medium have been analysed by fractional 
integration and differentiation~\citep[][and references therein]{zas}.
To do so, the fractal medium has had to be replaced 
by a continuous medium and the integrals on the network of the fractal 
medium has had to be approximated by fractional integrals~\citep{ren}.
The interpretation of fractional integration is connected with 
fractional mass dimension~\citep{mand}. Fractional integrals can 
be considered as integrals over fractional dimension space within 
a numerical factor~\citep{tara}. This numerical factor has been 
chosen suitably to get the right dimension of any physical parameter 
and to derive the standard expression in the continuum limit. The 
isotropic and homogeneous nature of dimensionality has also been 
incorporated properly. All of these will give a self-consistent 
description of the hydrodynamics in a fractal medium~\citep{roy07}.

Once the hydrodynamic equivalence has been established, it has then 
been a fairly easy exercise to model the steady fractal flow like a 
simple first-order autonomous dynamical system~\citep{stro,js99}. This 
has made it possible to gain an understanding of the critical aspects 
of the stationary phase portrait of the fractal flow, especially the 
behaviour of the transonic solution. The critical point in the phase
portrait has been shown to be a saddle point, and 
the transonic solution that has to pass through this point has been
shown to be infinitely sensitive to the choice of a boundary condition. 
While this bodes ill for the feasibility of transonicity itself within 
the stationary framework, all steady global solutions
(transonic or otherwise) have been found to be stable under a 
time-dependent linearised perturbation. An interesting fact that has
come to light is that the necessary mathematical conditions, which 
include an equation of motion for the dynamic perturbation and its 
relevant boundary 
conditions, to argue for the stability of the steady fractal
flows, have been found to be entirely identical to what has been 
reported earlier regarding the stability of continuous spherically
symmetric inflows~\citep{pso80,td92}. This similarity is fortunate
and armed with this knowledge, it can be safely claimed that fractal
flows are just as stable as continuous flows under the effects of
small linearised perturbations.

Having said this, one will still have to confront the fact that 
the time-dependent perturbative analysis has done nothing to indicate 
the primacy of the transonic solution, and that the steady
transonic inflow solution would not be possible without an infinite 
precision in prescribing a proper boundary condition. This obstacle
has, however, been overcome by taking into consideration explicit 
dynamics in the
flow system, and then evolving a physical flow through time, after 
having started with appropriate initial conditions. Transonicity 
becomes evident very soon, and it has been argued with a simplified
analytical model in the ``pressure-free" limit, that the guiding physical 
criterion to select the transonic solution is the one forwarded
by~\citet{bon52}, i.e. the transonic solution will be chosen because
it corresponds to a 
minimum energy configuration. While the ``pressure-free" limit does
not involve the fractal properties directly, it has been demonstrated 
through a numerical integration of the dynamic flow equations of the
fractal medium that transonicity is very much the favoured mode of 
infall in this case too. And the most salient result to have emerged 
from this numerical study has been that transonic features becomes 
more pronounced, as the medium is more like a fractal. 

It has already been mentioned that the perturbative treatment has 
been shown to give no clear-cut evidence to favour transonicity. Support 
for transonicity, however, has come indirectly from the perturbative
angle too. The equation of motion for the propagation of the linearised 
perturbation has been shown to have subtle similarities with an 
equation implying the metric of an acoustic black hole~\citep{vis98}. 
This hints at the fact that matter might cross the sonic horizon at
the greatest possible rate, i.e. transonically, just as matter has
to cross the event horizon of a black hole maximally. 

\section{The equations of the flow and its fixed points}
\label{sec2}

Considering the existence of a medium that has a fractal structure of 
mass dimensionality $D=3d$ (with $d<1$) embedded in a 3-dimensional space,
the mass enclosed in a sphere of radius $r$ can be written as~\citep{roy07} 
\begin{equation}
M_D = kr^D \sim \rho l_{\mathrm c}^3 
\left(\frac{r}{l_{\mathrm c}}\right)^{3d} , 
\label{md1}
\end{equation}
with $D$ referring to the dimension, $\rho$ to the constant density of 
the medium, and $l_{\mathrm c}$ to a characteristic inner length of the 
medium that can take an arbitrary value in the limit $d \longrightarrow 1$. 
This is the scale below which the medium will be continuous. The fractional 
integrals are computed as integrals over fractional dimension space 
within a numerical factor. The fractional infinitesimal length for 
a medium with isotropic mass dimension will, therefore, be 
given by~\citep{roy07} 
\begin{equation}
{\mathrm d}\overline{r}=\left(\frac{r}{l_{\mathrm c}}\right)^{d-1}
{\mathrm d}r , 
\label{dr}
\end{equation}
with the constant having been chosen to derive the standard expression in 
the limit $d\longrightarrow 1$. It is to be noted that the infinitesimal 
area and volume elements in this ``fractional continuous'' medium of 
mass dimension $D=3d$ will be different, and hence the mass enclosed 
in a sphere of radius $r$ for constant density $\rho$ will be~\citep{roy07}
\begin{equation}
M_D=\int_V\rho {\mathrm d}\overline{V}=\frac{4}{3}
\pi\rho \left(\frac{l_{\mathrm c}}{d}\right)^3 
\left(\frac{r}{l_{\mathrm c}}\right)^{3d} \sim r^D . 
\label{mdinteg}
\end{equation}

In this medium the inviscid Euler equation, describing the dynamics of 
the velocity field, $v$, can be expressed as~\citep{roy07}
\begin{equation}
\frac{\partial v}{\partial t}+v\frac{\partial v}{\partial r}
+\frac{1}{\rho}\frac{\partial p}{\partial r}+\phi^{\prime}(r)=0 , 
\label{euler}
\end{equation}
where $\phi(r)$ is the gravitational potential of the central 
accretor that drives the flow (with the prime denoting the spatial 
derivative of the potential). This is a local conservation law and, 
as it is to be expected, this has exactly the same form as that of 
the equation for the continuous medium. In the case of stellar accretion, 
the flow is driven by the Newtonian potential, $\phi = -GM/r$. On the
other hand, frequently in studies of black hole accretion, it becomes 
convenient to dispense 
completely with the rigour of general relativity, and instead make 
use of an ``effective'' pseudo-Newtonian potential that will imitate 
general relativistic effects in the Newtonian construct of space and 
time~\citep{pw80,nw91,abn96}. The choice of a particular form of the
potential will not affect the general arguments overmuch. 

The pressure, $p$, is related to the local density, $\rho$, 
through a polytropic 
equation of state $p=K \rho^\gamma$, in which $K$ is a constant, and 
$\gamma$ is the polytropic exponent, whose admissible range is given 
by $1< \gamma < 5/3$. This range is restricted by the isothermal 
and the adiabatic limits, respectively~\citep{sc39}. The evolution
of $\rho$ is described by the equation of continuity~\citep{roy07}, 
\begin{equation}
\frac{\partial \rho}{\partial t}+\frac{1}{r^\alpha}
\frac{\partial}{\partial r}\left(\rho v r^\alpha \right)=0 , 
\label{con}
\end{equation}
in which $\alpha = 3d -1$. 

The flow system will, therefore, be specified by equations~(\ref{euler}) 
and~(\ref{con}), along with the polytropic equation of state. Of 
particular interest are the steady state solutions for the case where 
the fractal medium is at rest at a large distance from the accretor. 
Since transonic flows are of primary concern here, it is required that 
the static flow should evolve from $v \longrightarrow 0$ as 
$r \longrightarrow {\infty}$ (the outer boundary condition) to $v> a(r)$ 
for small $r$, where $a$ is the speed of sound, given by 
$a^2 = \partial p/ \partial \rho = \gamma K \rho ^{\gamma - 1}$. The 
stationary state implies 
$\partial v/ \partial t = \partial \rho /\partial t = 0$.  
Consequently, equations~(\ref{euler}) and~(\ref{con}) will be reduced 
to their steady state forms as  
\begin{equation}
v\frac{{\mathrm d}v}{{\mathrm d}r}
+\frac{1}{\rho}\frac{{\mathrm d}p}{{\mathrm d}r}+\phi^{\prime}(r)=0
\label{stateuler}
\end{equation}
and
\begin{equation}
\frac{{\mathrm d}}{{\mathrm d}r}\left(\rho v r^\alpha \right) =0 , 
\label{statcon}
\end{equation}
respectively. It is easily seen that equations~(\ref{stateuler}) 
and~(\ref{statcon}) remain invariant under the transformation 
$v \longrightarrow -v$, i.e. the mathematical problem for inflows 
($v<0$) and outflows ($v>0$) is identical in the stationary 
state~\citep{arc}.

\begin{figure}
\begin{center}
\includegraphics[scale=0.65, angle=0.0]{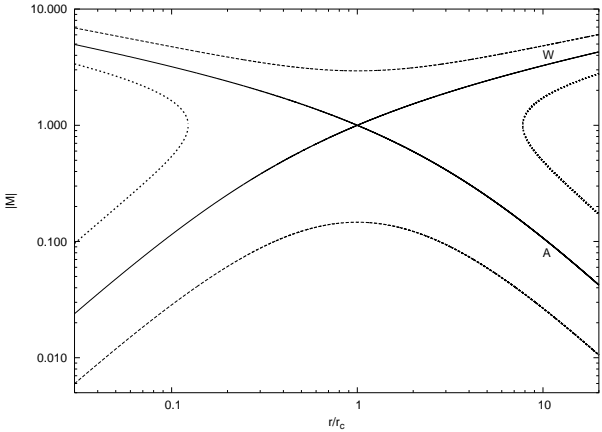}
\caption{\label{f1} \small{Integral solutions of the stationary 
fractal flow, driven by the Newtonian potential, for $D=2.55$ 
and $\gamma=1.4$. The 
continuous curves, $\mathrm A$ and $\mathrm W$ represent ``accretion" 
and ``wind", respectively. The fixed point is at $r=r_{\mathrm c}$ 
and $\vert {\mathrm M} \vert =1$. This point is a saddle point.}}
\end{center}
\end{figure}

It is in principle possible to eliminate either $v$ or $\rho$ and solve 
for the other variable as a function of $r$. However, adopting a slightly 
different approach, it is possible to recast equations~(\ref{stateuler}) 
and~(\ref{statcon}) in a combined form as
\begin{equation}
\frac{\mathrm d}{{\mathrm d}r} \left(v^2 \right) = \frac{2 v^2}{r} 
\left[\frac{\alpha a^2 - r \phi^{\prime}(r)}{v^2 - a^2} \right] , 
\label{dvdr}
\end{equation}
whose integral solutions have been shown in Fig.~\ref{f1}, with 
$\phi(r) = - GM/r$, and with the vertical axis scaled by the 
absolute value of the Mach
number, $\mathrm{M} = v/a$. The two continuous curves labelled ``A" and 
``W" refer to accretion and wind, respectively. The critical points 
in the flow will be derived from the standard requirement that the 
flow solutions will have a finite gradient when they will cross the 
sonic horizon (where the bulk flow velocity exactly matches the speed
of acoustic propagation), which will mean that both the numerator 
and the denominator of equation~(\ref{dvdr}) will have to vanish 
simultaneously and non-trivially~\citep{skc90,skc96}. This can only 
happen when 
\begin{equation}
v_{\mathrm c}^2 = a_{\mathrm{c}}^2 = 
\frac{r_{\mathrm c}\phi^{\prime}(r_{\mathrm c})}{\alpha} , 
\label{critcon1}
\end{equation}
which gives the critical point (or the sonic point in this particular
case) conditions, with the subscript 
``$\mathrm c$" labelling the critical point values.

It is not a difficult exercise to integrate equation~(\ref{stateuler}) 
and then transform the variable $\rho$ in it to $a$ with the help of 
the equation of state. This, with the critical point conditions as 
given by equation~(\ref{critcon1}), will give a relation for fixing 
the critical point coordinates in terms of the flow parameters, 
$\mathcal E$ (which is actually Bernoulli's constant), $\alpha$ 
and $\gamma$ as 
\begin{equation}
\left(\frac{\gamma +1}{\gamma -1}\right) \frac{r_{\mathrm c}\phi^{\prime}
(r_{\mathrm c})}{2\alpha} + \phi(r_{\mathrm c}) = {\mathcal E} . 
\label{eulerfix}
\end{equation}
The form of $\phi(r)$ will obviously determine the number of the
critical points, and for the Newtonian potential only one root of
$r_{\mathrm c}$ will be obtained from equation~(\ref{eulerfix}). 
This root, for $\mathcal E$ fixed by the outer boundary condition
of the transonic inflow solution, will be given as 
\begin{equation}
r_{\mathrm c} = \left[\frac{\left(\gamma+1\right)-2\alpha
\left(\gamma-1\right)}{2\alpha}\right] \frac{GM}{a_\infty^2} , 
\label{rcrit}
\end{equation}
with $a_\infty$ being the speed of sound at the outer boundary of the
flow~\citep{skc90,skc96}, where the influence of gravity is negligibly 
weak.

It should be worth mentioning here that although with the choice of 
a pseudo-Newtonian potential multiple roots for $r_{\mathrm c}$
would result, practically speaking only one of these roots would
be a physically meaningful critical point, through which an integral
solution may pass. For spherically symmetric flows in the general 
relativistic framework, this issue has been discussed by~\citet{mrd07}. 

\section{The flow as an autonomous dynamical system}
\label{sec3}

So far the flow variables have been ascertained only at the critical
points. Since the flow equations are in general nonlinear differential
equations, short of carrying out a numerical integration, there is 
no completely rigorous analytical prescription for solving these 
differential equations to determine the global nature of the flow 
variables.
Nevertheless, some analytical headway could be made after all by taking
advantage of the fact that equation~(\ref{dvdr}), which gives a
complete description of the $r$ --- $v^2$ phase portrait of the flow,
is an autonomous first-order differential equation, and as such, could
easily be recast into the mathematical form ${\dot{x}} = X(x,y)$ and
${\dot{y}} = Y(x,y)$, which is that of the very familiar coupled
first-order dynamical system~\citep{stro,js99}.
Quite frequently for any nonlinear physical system, a linearised
analytical study of the properties of the fixed points of a 
first-order dynamical system, affords a
robust platform for carrying out an investigation to understand
the global behaviour of integral solutions in the phase portrait.

And so it is that to investigate the nature of the critical point, 
equation~(\ref{dvdr}) will have to be decomposed in terms of a 
mathematical parameter, $\tau$, to read as
\begin{eqnarray}
\frac{\mathrm d}{{\mathrm d}\tau} \left(v^2\right) &=& 
2 v^2 \left[\alpha a^2 - r \phi^{\prime}(r)\right] 
\nonumber \\
\frac{{\mathrm d}r}{{\mathrm d}\tau} &=& r\left(v^2 - a^2\right) , 
\label{dynsys}
\end{eqnarray}
in both of which the parameter $\tau$ does not make an explicit 
appearance in the right hand side, something of an especial advantage 
that derives from working with autonomous systems. This kind of 
parametrization is quite common in fluid dynamics~\citep{bdp93}, 
and in accretion studies especially, this approach has been made 
before~\citep{rb02,ap03,crd06,mrd07,gkrd07}. Some earlier works in 
accretion 
had also made use of the general mathematical aspects of this 
approach~\citep{mkfo84,mc86,ak89}. A further point that has to 
be noted is that the function $a^2$ in the right hand side of 
equation~(\ref{dynsys}) can be expressed entirely in terms of 
$v^2$ and $r$, with the help of the equation of state and 
equation~(\ref{statcon}). This will exactly satisfy the 
criterion of a first-order autonomous dynamical system.

The next task would be to make a linearised approximation about 
the fixed point coordinates and extract a linear dynamical system 
out of equations~(\ref{dynsys}). This will give a direct way to 
establish the nature of the critical points (or fixed points). 
Expanding about the fixed point values, a perturbation of the kind 
$v^2=v_{\mathrm c}^2 +\delta v^2=v_{\mathrm c}^2(1 +\epsilon_1)$ 
and $r=r_{\mathrm c} +\delta r=r_{\mathrm c}(1 +\epsilon_2)$ 
can be applied. Using the continuity equation and the equation 
of state, this perturbation scheme, when linearised, will also 
give $\delta a/a^2_{\mathrm c}=-\left(\gamma -1\right)
\left(\epsilon_1+2\alpha\epsilon_2\right)/2$. Applying this 
perturbative expansion on equation~(\ref{dynsys}), and linearising 
in $\epsilon_1$ and $\epsilon_2$ will give,
\begin{eqnarray}
\frac{{\mathrm d}\epsilon_1}{{\mathrm d}\tau}&=&\alpha v_{\mathrm c}^2 
\Bigg\{-\left(\gamma -1\right)\epsilon_1 \nonumber \\
& &-2 \left[\alpha\gamma -\alpha 
+1 +\frac{\phi^{\prime\prime}(r_{\mathrm c}) r_{\mathrm c}}
{\phi^{\prime}(r_{\mathrm c})} \right] 
\epsilon_2 \Bigg\} \nonumber \\
\frac{{\mathrm d}\epsilon_2}{{\mathrm d} \tau}&=&\alpha 
v_{\mathrm c}^2 \left[\left(\frac{\gamma +1}{2\alpha} \right)
\epsilon_1 +\left(\gamma-1 \right)\epsilon_2 \right] .
\label{lindyn}
\end{eqnarray}
Using solutions of the type $\epsilon_1 \sim \exp(\Omega \tau)$ 
and $\epsilon_2 \sim \exp(\Omega \tau)$ in equations~(\ref{lindyn}), 
the eigenvalues of the stability matrix associated with the critical 
points will be derived as 
\begin{equation}
\Omega^2 = \alpha a_{\mathrm c}^4 \left[\left(2\alpha-1\right)
-\gamma\left(2\alpha+1\right)-\left(\gamma+1\right)r_{\mathrm c}
\frac{\phi^{\prime\prime}(r_{\mathrm c})}{\phi^{\prime}(r_{\mathrm c})}
\right] 
\label{eigen}
\end{equation}
with $a_{\mathrm c}$ itself being a function of $r_{\mathrm c}$, as
given by equation~(\ref{critcon1}). 

Once the position of a critical point, $r_{\mathrm c}$, has become 
known from equation~(\ref{eulerfix}), it is then quite easy to determine 
the nature of that critical point by using $r_{\mathrm c}$ in 
equation~(\ref{eigen}). Since $r_{\mathrm c}$ is a function of 
$\mathcal E$ and $\gamma$, it effectively implies that $\Omega^2$ can, 
in principle, be regarded as a function of the flow parameters. From 
the form of $\Omega^2$ in equation~(\ref{eigen}), a generic conclusion 
that can be immediately drawn is that any critical point, as it may
be expected for a conservative system, will be either a saddle point 
(for $\Omega^2 >0$) or a centre-type point (for $\Omega^2 <0$). For the 
particular case of the Newtonian potential, $\phi=-GM/r$, the eigenvalues
of the stability matrix will be given by 
\begin{equation}
\Omega_{\mathrm N}^2 =\alpha a_{\mathrm c}^4 
\left[\left(2\alpha+1\right)-\gamma\left(2\alpha-1\right)\right] .
\label{neigen}
\end{equation}
For the values of $\gamma$ and $\alpha$ lying in the range of physical 
interest, it can always be shown that $\Omega_{\mathrm N}^2 > 0$. Hence 
in this case 
the critical point is a saddle point, and the curves which have been 
labelled ``accretion" and ``wind" in Fig.~\ref{f1} are in fact separatrices 
of a dynamical system, rather than physical solutions. 

The understand the full import of this line of reasoning, what has to
be borne in mind is that saddle points are inherently unstable, and to 
make a solution pass through such a point, after starting from an outer 
boundary condition, will entail an infinitely precise fine-tuning of that
boundary condition~\citep{rb02}. This can be demonstrated through simple
arguments. Going back to equations~(\ref{lindyn}), the coupled set of 
linear differential equations in $\epsilon_1$ and $\epsilon_2$ can be 
set down as
\begin{equation}
\label{ratio}
\frac{{\mathrm d} \epsilon_1}{{\mathrm d} \epsilon_2} 
= \frac{{\mathrm d} \epsilon_1/{\mathrm d} \tau}
{{\mathrm d} \epsilon_2/{\mathrm d} \tau} =
\frac{{\mathcal Q}_1 \epsilon_1 + {\mathcal Q}_2 \epsilon_2}
{{\mathcal Q}_3 \epsilon_1 + {\mathcal Q}_4 \epsilon_2} , 
\end{equation}
in which the constant coefficients ${\mathcal Q}_1$, ${\mathcal Q}_2$,
${\mathcal Q}_3$ and ${\mathcal Q}_4$ are to be determined simply by
an inspection of equations~(\ref{lindyn}). It is also to be easily seen 
that ${\mathcal Q}_1 = - {\mathcal Q}_4$. This makes the integration 
of equation~(\ref{ratio}) a straightforward exercise, and it  yields
\begin{equation}
\label{hyper}
{\mathcal Q}_2 \epsilon_2^2 +2{\mathcal Q}_1 \epsilon_1 \epsilon_2
- {\mathcal Q}_3 \epsilon_1^2 + {\mathcal C} = 0
\end{equation}
with ${\mathcal C}$ being an integration constant. Generally speaking
equation~(\ref{hyper}) is the equation of a conic section in the
$\epsilon_2$ --- $\epsilon_1$ plane. If the origin of this plane were
to be considered to have been shifted to the saddle point, then the
condition for solutions passing through the origin,
i.e. $\epsilon_1 = \epsilon_2 = 0$, would be ${\mathcal C} = 0$, 
which would reduce equation~(\ref{hyper}) to a pair of straight
lines intersecting each other through the origin itself. All other
solutions in the vicinity of the origin will, therefore, be hyperbolic
in nature, a fact that is given by the condition
$({\mathcal Q}_1^2 + {\mathcal Q}_2 {\mathcal Q}_3) > 0$.
For the case of $\phi = -GM/r$, this contention can be verified
completely analytically, and this shows that even a very minute
deviation from a precise boundary condition for transonicity
(i.e. a boundary condition
that will generate solutions to pass {\em only} through the origin,
$\epsilon_1 = \epsilon_2 = 0$) will take the stationary solution far
away from a transonic state. This extreme sensitivity of transonic
solutions to boundary conditions is entirely in keeping with the
nature of saddle points. It may be imagined that in a proper
astrophysical system such precise fulfillment of a boundary
condition will make the transonic solution well-nigh physically
non-realisable. Indeed, this difficulty, for any kind of accreting
system, is readily appreciated by anyone trying to carry out a numerical
integration of equation~(\ref{dvdr}) to generate the transonic solutions, 
which can only be obtained when the numerics is first biased in favour 
of transonicity by using the saddle point condition itself as the
boundary condition for numerical integration.

Apart from this, there is also a mathematical aspect of the physical 
non-realisability of transonic solutions. 
Using the condition ${\mathcal C} = 0$ will make it easy
to express $\epsilon_1$ in terms of $\epsilon_2$ and vice versa. 
Going back to the set of linear equations given by equations~(\ref{lindyn}) 
and choosing the second one of the two equations (the choice of 
the first would also have led to the same result), one gets
\begin{equation}
\label{aar}
\frac{{\mathrm d} \epsilon_2}{{\mathrm d} \tau}= \pm
\left(\sqrt{{\mathcal Q}_1^2 + {\mathcal Q}_2 {\mathcal Q}_3}\right) 
\,\,\epsilon_2 , 
\end{equation}
which can be integrated for both the roots from an arbitrary
initial value of $\epsilon_2 = \epsilon_2^{\star}$ lying anywhere
on the transonic solution, to a point $\epsilon_2 = \Delta$, with 
$\Delta$ being very close to the critical point given by 
$\epsilon_2 =0$. Using the equivalence that 
$\Omega^2 = {\mathcal Q}_1^2 + {\mathcal Q}_2 {\mathcal Q}_3$, 
it can be shown that
\begin{equation}
\label{tau}
\tau = \pm \frac{1}{\Omega}\int_{\epsilon_2^{\star}}^{\Delta} 
\frac{{\mathrm d} \epsilon_2}{\epsilon_2} 
= \pm \frac{1}{\Omega}
\ln \Bigg{\vert} \frac{\Delta}{\epsilon_2^{\star}} \Bigg{\vert} , 
\end{equation}
from which it is easy to see that 
$\vert\tau\vert\longrightarrow\infty$
for $\Delta \longrightarrow 0$. This implies
that the critical point may be reached along either of the
separatrices, only after $\vert \tau \vert$ has become
infinitely large. This divergence of the parameter $\tau$
indicates that in the stationary regime, solutions passing
through the saddle point are not actual solutions, but
separatrices of various classes of solutions~\citep{stro,js99}. This fact,
coupled with the sensitivity of the stationary transonic solutions
to the choice of an outer boundary condition, makes
their feasibility a seriously questionable matter.

\section{A time-dependent perturbative approach}
\label{sec4}

It has been demonstrated in the foregoing section that the steady 
transonic accretion solution is unstable under infinitesimal deviations
from the precise outer boundary condition needed to generate the 
solution. In 
the astrophysical context, such precision is quite impossible,
and, therefore, the very feasibility of transonicity becomes a matter
of grave doubt. This difficulty can, however, be dispelled if one is
mindful of the fact that the real astrophysical flow is not static
but dynamic in character, i.e. it will have an explicit dependence
on time. This, of course, will mean that the time-dependent terms 
involving both the velocity and the density fields in 
equations~(\ref{euler}) and~(\ref{con}) will have to be retained. 

While full-fledged time-dependence of the flow variables will 
undoubtedly reveal many new interesting mathematical facets (all of
them involving the mathematics of partial differential equations) 
of the physical problem, 
it would still be worthwhile to go back to studying the properties
of the background stationary flow under the influence of a linearised
perturbative effect. As a preliminary exercise in accounting for
explicit time-dependence, this will at least shed some light on the 
global stability of the flow solutions. 

To that end, it will first be necessary to define, closely following a 
perturbative procedure prescribed by~\citet{pso80} and~\citet{td92}, 
a new physical variable $f=\rho v r^\alpha$. It is quite obvious from 
the form of equation~(\ref{con}) that the stationary value of $f$ will 
be a global constant, $f_0$, which can be closely identified with the 
matter flux rate. A perturbation prescription of the form 
$v(r,t) = v_0(r) + v^{\prime}(r,t)$ and 
$\rho (r,t) = \rho_0 (r) + \rho^{\prime}(r,t)$, will give, 
on linearising in the primed quantities,
\begin{equation}
f^{\prime} = f_0 \left(\frac{\rho^{\prime}}{\rho_0} 
+ \frac{v^{\prime}}{v_0} \right) , 
\label{effprime}
\end{equation}
with the subscript ``$0$" denoting stationary values in all cases. 
From equation~(\ref{con}), it then becomes possible to set down the 
density fluctuations, $\rho^{\prime}$, in terms of $f^{\prime}$ as 
\begin{equation}
\frac{\partial \rho^{\prime}}{\partial t} +\frac{v_0 \rho_0}{f_0} 
\left(\frac{\partial f^{\prime}}{\partial r}\right)=0 .
\label{flucden}
\end{equation}
Combining equations~(\ref{effprime}) and~(\ref{flucden}) will then 
render the velocity fluctuations as 
\begin{equation}
\frac{\partial v^{\prime}}{\partial t}= 
\frac{v_0}{f_0}\left(\frac{\partial f^{\prime}}{\partial t}
+ v_0 \frac{\partial f^{\prime}}{\partial r}\right)
\label{flucvel}
\end{equation}
which, upon a further partial differentiation with respect to time, 
will give 
\begin{equation}
\frac{{\partial}^2 v^{\prime}}{\partial t^2}=\frac{\partial}{\partial t} 
\left[\frac{v_0}{f_0} \left(\frac{\partial f^{\prime}}{\partial t}\right) 
\right]+\frac{\partial}{\partial t} \left[ \frac{v_0^2}{f_0} 
\left(\frac{\partial f^{\prime}}{\partial r}\right)\right] . 
\label{flucvelder2}
\end{equation}

From equation~(\ref{euler}) the linearised fluctuating part could be 
extracted as 
\begin{equation}
\frac{\partial v^{\prime}}{\partial t}+ \frac{\partial}{\partial r}
\left( v_0 v^{\prime} + a_0^2 
\frac{\rho^{\prime}}{\rho_0}\right) =0
\label{fluceuler}
\end{equation}
with $a_0$ being the speed of sound in the steady state. Differentiating 
equation~(\ref{fluceuler}) partially with respect to $t$, and making 
use of equations~(\ref{flucden}),~(\ref{flucvel}) and~(\ref{flucvelder2}) 
to substitute for all the first and second-order derivatives of 
$v^{\prime}$ and $\rho^{\prime}$, will deliver the result 
\begin{displaymath}
\frac{\partial}{\partial t} \left[\frac{v_0}{f_0}
\left( \frac{\partial f^{\prime}}{\partial t}\right)\right]
+ \frac{\partial}{\partial t} \left[\frac{v_0^2}{f_0}
\left( \frac{\partial f^{\prime}}{\partial r}\right)\right]
+ \frac{\partial}{\partial r} \left[\frac{v_0^2}{f_0}
\left( \frac{\partial f^{\prime}}{\partial t}\right)\right] 
\end{displaymath}
\begin{equation}
\qquad \qquad \qquad \qquad \qquad + \frac{\partial}{\partial r}
\left[\frac{v_0}{f_0}\left(v_0^2 - a_0^2 \right)
\frac{\partial f^{\prime}}{\partial r}\right] = 0 .
\label{interm}
\end{equation}
A little readjustment of terms in equation~(\ref{interm}) will finally 
give an equation of motion for the perturbation as  
\begin{equation}
\frac{{\partial}^2 f^{\prime}}{\partial t^2}+2\frac{\partial}{\partial r}
\left(v_0 \frac{\partial f^{\prime}}{\partial t}\right)+\frac{1}{v_0}
\frac{\partial}{\partial r}\left[ v_0 \left(v_0^2- a_0^2\right)
\frac{\partial f^{\prime}}{\partial r}\right] = 0 , 
\label{tpert}
\end{equation}
which is an expression that is exactly the same as what can be derived
upon perturbing the stationary solutions of spherically symmetric 
inflows in a continuous medium~\citep{pso80,td92}. Another aspect of
equation~(\ref{tpert}) is that its form has no explicit dependence on 
the potential --- Newtonian or pseudo-Newtonian --- that is driving the
flow. This is entirely to be expected, because the potential, being 
independent of time, will only lend its direct presence to the stationary
background flow. Arguments regarding stability will, therefore, be more
dependent on the boundary conditions of the steady flow. 
As the form of the equation of motion for the linearised perturbation 
remains unchanged even for a flow in a fractal medium, and as the 
physical boundary conditions are also not altered in this case, the 
general conclusions reached by both~\citet{pso80} and~\citet{td92}
regarding flows in a continuous medium, will carry over here, and it 
can be safely claimed that under all reasonable boundary conditions, 
both the transonic and subsonic solutions will be stable. 

While this does nothing to cause any immediate worry, it also does 
not reveal anything in particular either about the physical feasibility 
of any solution
from a perturbative point of view. This is in keeping with the 
conventional wisdom about spherically symmetric flows~\citep{gar79} 
that the natural
preference of the system for any particular solution --- especially 
the transonic solution --- cannot be justified by a linear stability
analysis, but by the more fundamental arguments forwarded 
by~\citet{bon52}. 

For all that, some positive hint about the primary status of the
transonic solution can actually be derived if the whole issue of
a linear stability analysis is viewed from a different perspective. 
It is known that there is a close one-to-one correspondence between 
certain features of black hole physics and the physics of supersonic 
acoustic flows \citep{vis98}. In this very context, a compact 
rendering of equation~(\ref{interm}) can be obtained as 
\begin{equation}
\partial_\mu \left({\mathrm{f}}^{\mu \nu} 
\partial_\nu f^{\prime}\right) = 0 , 
\label{compact}
\end{equation}
in which the Greek indices are made to run from $0$ to $1$, with the 
identification that $0$ stands for $t$, and $1$ stands for $r$. An 
inspection of the terms in the left hand side of equation~(\ref{interm}) 
will then allow for constructing the symmetric matrix
\begin{equation}
{\mathrm{f}}^{\mu\nu}=\frac{v_0}{f_0}
\pmatrix
{1 & v_0 \cr 
v_0 & v_0^2 - a_0^2} . 
\label{matrix}
\end{equation}
Now in Lorentzian geometry the d'Alembertian for a scalar field in 
curved space is given in 
terms of the metric ${\mathrm{g}}_{\mu \nu}$ by~\citep{vis98}
\begin{equation}
\Delta \psi\equiv \frac{1}{\sqrt{-\mathrm{g}}}
\partial_\mu\left({\sqrt{-\mathrm{g}}}\, 
{\mathrm{g}}^{\mu\nu}\partial_\nu\psi\right)
\label{alem}
\end{equation}
with $\mathrm{g}^{\mu\nu}$ being the inverse of the matrix implied 
by ${\mathrm{g}}_{\mu\nu}$. Comparing equation~(\ref{compact}) 
with equation~(\ref{alem}), 
it would be tempting to look for an exact equivalence between
${\mathrm f}^{\mu\nu }$ and $\sqrt{-\mathrm g}\, {\mathrm g}^{\mu\nu}$.
This, however, cannot be done in a general sense. What can be
appreciated, nevertheless, is that equation~(\ref{compact}) gives 
a relation for $f^{\prime}$ which is of the type given by 
equation~(\ref{alem}). The metrical part of equation~(\ref{compact}), 
as given by equation~(\ref{matrix}), may then be
extracted, and its inverse will incorporate the notion of a sonic
horizon of an acoustic black hole when $v_0^2 = a_0^2$.
This point of view does not make for a perfect acoustic analogue model,
but it has some similar features to the metric of a wave equation for a
scalar field in curved space-time, obtained through a
somewhat different approach, in which the velocity of an an irrotational,
inviscid and barotropic fluid flow is first represented as the gradient
of a scalar function $\psi$, i.e. ${\bf v}= -{\bf{\nabla}}\psi$, and
then a perturbation is imposed on this scalar function~\citep{vis98}.

The foregoing discussion indicates that the physics of supersonic 
acoustic flows closely corresponds to many features of black hole 
physics. This closeness of form is very intriguing. For a black hole, 
infalling matter crosses the event horizon maximally, i.e. at the 
greatest possible speed. By analogy the same thing may be said of 
matter crossing the sonic horizon of a spherically symmetric fluid 
flow, falling on to a point sink. That this fact can be appreciated 
for the spherically symmetric accretion problem, through a perturbative 
result as given by equation~(\ref{interm}), is quite remarkable. This 
is because it is universally recognised that that no insight into the 
special status of any inflow solution may possibly be derived solely 
through a perturbative technique~\citep{gar79}. It is the 
transonic solution that crosses the sonic horizon at the greatest 
possible rate~\citep{bon52}, 
and the similarity of form between equations~(\ref{interm}) 
and~(\ref{alem}) may very well be indicative of the primacy of the 
transonic solution. If such an insight were truly to be had with the 
help of the perturbation equation, then the perturbative linear 
stability analysis might not have been carried out in vain after all.

\section{Dynamic evolution towards transonicity}
\label{sec5}

Much more direct and robust evidence in favour of transonicity could 
be obtained if the
accreting system were to be made to evolve through time, as opposed
to making it suffer small linearised perturbations in time. Having 
said this, it must also be stressed that equations~(\ref{euler}) 
and~(\ref{con}), which govern the temporal evolution of the flow, 
are not amenable to a ready mathematical analysis; indeed, in the matter
of incorporating both the dynamic and the pressure effects in 
the equations, short of a direct numerical treatment, the mathematical 
problem, is very aptly described as ``insuperable"~\citep{bon52}.
Therefore, to have a preliminary 
appreciation of the governing mechanism that underlies any possible 
selection of a transonic flow, it should be necessary to adopt
some simplifications. This will pave the way for a more complete
physical understanding of the evolutionary properties of the flow. 

The evolutionary dynamics is, therefore, to be studied first in the 
regime of what is understood to be the ``pressureless" motion of a fluid 
in a gravitational field~\citep{shu}, as opposed to dropping the 
dynamic effects to study a much simplified 
stationary picture~\citep{bon52}. Simplification of 
the mathematical equations, however, is not the only justification
for such a prescription. A greater justification lies in the fact
that the result delivered is in conformity with, what~\citet{gar79} 
calls ``the more fundamental arguments" of~\citet{bon52}; that it 
is the criterion of minimum total energy associated with a solution,
that will accord it a principal status over all the others. 

An immediate consequence of adopting dynamic equations is that
the invariance of the stationary solutions under the transformation
$v \longrightarrow -v$, is lost. As a result, one will have to 
separately consider either the inflows $(v<0)$ or the outflows $(v>0)$, 
a choice that has to be imposed upon the system at $t=0$. Euler's 
equation, tailored according to the simplified requirements of a 
``pressureless" field, is rendered as
\begin{equation}
\frac{\partial v}{\partial t} + v \frac{\partial v}{\partial r} 
+ \frac{GM}{r^2} = 0
\label{presfree}
\end{equation}
which can be solved by the method of characteristics~\citep{ld97}. 
The characteristic solutions are obtained from
\begin{equation}
\frac{{\mathrm d}t}{1}= \frac{{\mathrm d}r}{v} 
= \frac{{\mathrm d}v}{-GM/r^2} . 
\label{charcur}
\end{equation}
First solving the ${\mathrm d}v/{\mathrm d}r$ equation will give
\begin{equation}
\frac{v^2}{2}- \frac{GM}{r} = \frac{c^2}{2} 
\label{dvdrchar}
\end{equation}
in which $c$ is an integration constant that derives from the spatial 
part of the characteristic equation. This result is to be used to solve 
the ${\mathrm d}r/{\mathrm d}t$ equation from equation~(\ref{charcur}), 
which will finally lead to 
\begin{equation}
\frac{2vr}{c r_{\mathrm s}} -\ln r -
\ln \left(\frac{v}{c}+1 \right)^2
- \frac{2ct}{r_{\mathrm s}} = {\tilde c}  
\label{drdt}
\end{equation}
with $\tilde{c}$ being another integration constant, and $r_{\mathrm s}$ 
being a length scale in the system, defined as $r_{\mathrm s}=2GM/c^2$.

A general solution of equation~(\ref{charcur}) is given by the 
condition, $f({\tilde c})= c^2/2$, with $f$ being an arbitrary 
function, whose form is to be determined from the initial condition. 
The general solution can, therefore, be set down as 
\begin{equation}
\frac{v^2}{2}-\frac{GM}{r}=f\left[\frac{2vr}{c{r_{\mathrm s}}} -\ln r 
-\ln \left(\frac{v}{c}+1\right)^2-\frac{2ct}{r_{\mathrm s}}\right] , 
\label{gensol2}
\end{equation}
to determine whose particular form, the initial condition that will
have to be used is $v=0$ at $t=0$ for all $r$. This will lead to  
\begin{equation}
\frac{v^2}{2}-\frac{GM}{r}= -\frac{GM}{r}\left(\frac{v}{c}+1 
\right)^{-2}\exp\left(\frac{2vr}{cr_{\mathrm s}} 
-\frac{2ct}{r_{\mathrm s}}\right) 
\label{partform}
\end{equation}
from which it is easy to see that for $t \longrightarrow \infty$, 
what is approached is the stationary solution,  
\begin{equation}
\frac{v^2}{2} - \frac{GM}{r} = 0 . 
\label{statsol}
\end{equation}
Corresponding to the given initial condition, this is evidently
the stationary solution associated with the lowest possible total 
energy, and the temporal evolution selects this solution from among
all the others. The whole picture could be conceived of as one in 
which a system with a uniform velocity distribution $v=0$ everywhere, 
suddenly has a gravity mechanism switched on in its midst at $t=0$. 
This induces a potential $-GM/r$ at all points in space. The system 
then starts evolving to restore itself to another stationary state, 
with the velocity increasing according to equation~(\ref{partform}), 
so that for $t \longrightarrow \infty$, the total energy at all 
points, $E=(v^2/2)-(GM/r)=0$, remains the same as at $t=0$.

\begin{figure}
\begin{center}
\includegraphics[scale=0.65, angle=0.0]{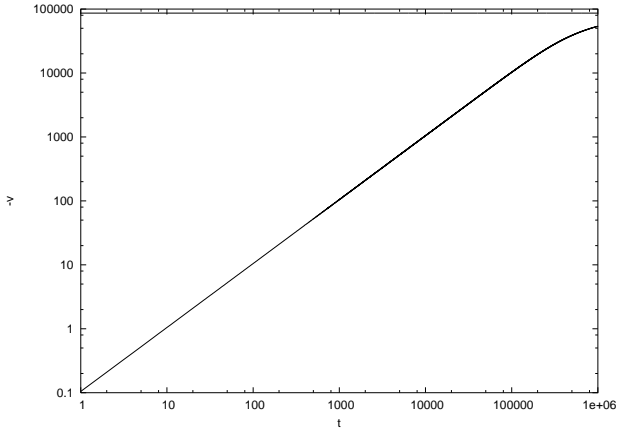}
\caption{\label{f2} \small{For the temporal evolution of velocity,
the slope of this logarithmic plot shows that in the early stages 
of the evolution, $-v$ varies linearly with $t$. Deviation from  
linear growth sets in later. The horizontal 
line on top shows the terminal value of the velocity field, whose 
evolution is being followed at a fixed length scale.}}
\end{center}
\end{figure}

\begin{figure}
\begin{center}
\includegraphics[scale=0.65, angle=0.0]{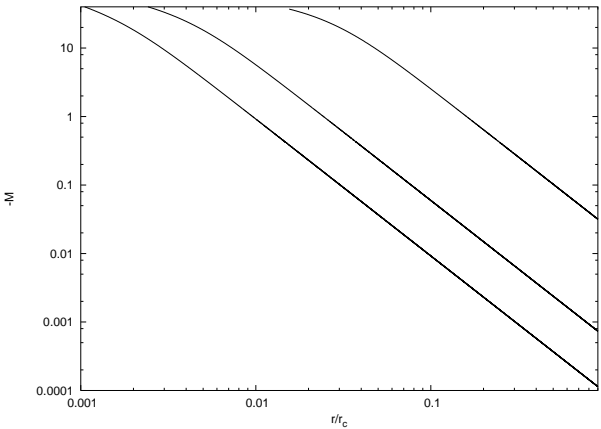}
\caption{\label{f3} \small{The $\log$-$\log$ plot of the Mach number 
versus the radial distance, with the latter being scaled by the sonic 
radius, $r_\mathrm{c}$, after $4000$ seconds of evolution. 
Going from left to right each curve corresponds to $d=0.7$, $0.85$
and $1.0$, respectively. The slope of the curves indicates a power
law behaviour for the evolution on intermediate length scales. All
the curves approach a saturation velocity scale closer to the accretor.}}
\end{center}
\end{figure}

\begin{figure}
\begin{center}
\includegraphics[scale=0.65, angle=0.0]{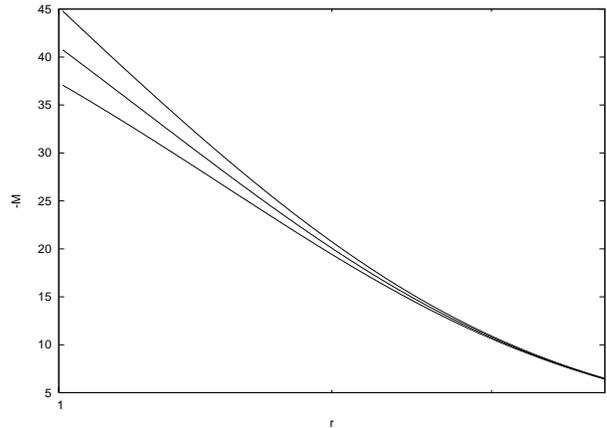}
\caption{\label{f4} \small{Velocity field, scaled as the Mach number, 
$\mathrm M$, after $4000$ seconds, for various values of $d$. Moving 
from the bottom to the top, successive solutions 
have been shown for $d=1.0$, $0.85$ and $0.7$, respectively. The 
radial distance has been normalised with respect to $r_\odot$, and
plotted logarithmically over a length scale of $r_\odot$ to 
$4r_\odot$.}}
\end{center}
\end{figure}

This contention has been borne out by a numerical integration of 
equation~(\ref{presfree}) by the finite differencing technique. The 
mass of the accretor has been chosen to be $M_\odot$, while its 
radius is $r_\odot$. The evolution through time has been followed 
at a fixed length scale of $51 r_\odot$. The result of the numerical
evolution of the velocity field, $-v$ (for inflows $v$ is actually
negative), through time, $t$, has been plotted in 
Fig.~\ref{f2}. The limiting value of the velocity, as 
the evolution progresses towards the long-time limit, is evidently
$\sqrt{2GM/r}$ (with $M= M_\odot$ and $r =51 r_\odot$), 
as equation~(\ref{statsol}) would suggest. 
This is what the plot in Fig.~\ref{f2} shows, as $-v$ approaches
its terminal value for $t \longrightarrow \infty$. The slope of 
this logarithmic plot also indicates that in the early
stages of the evolution there is a linear growth of the velocity
field through time, but on later times, conspicuous deviation from 
linearity sets in.  

The argument presented above, with the effects of pressure taken
into account, can now be extended to understand the
dynamic selection of the transonic solution. The inclusion of the
pressure term in the dynamic equation, fixes the total energy of
the system accordingly at $t=0$. A physically realistic initial 
condition should be that $v=0$ at $t=0$, for all $r$, while $\rho$ 
has some uniform value. The temporal evolution of the accreting 
system would then non-perturbatively select the transonic trajectory, 
as it is this solution with which is associated the least possible 
energy configuration. This argument is in conformity with the assertion 
made by~\citet{bon52} that it is the criterion of minimum total energy 
that should make a particular solution (the transonic solution in this 
case) preferred to all the others. However, this selection mechanism 
is effective only through the temporal evolution of the flow.

To test this contention a numerical study has been carried out once 
again using finite differencing, but this time using both the dynamic
equations for the velocity and the density fields, as given by 
equations~(\ref{euler}) and (\ref{con}). The accretor has been chosen 
to have a mass, $M_\odot$, and radius, $r_\odot$. The ``ambient" 
conditions are $a_\infty = 10 \ {\rm{km}}~{\rm{s}}^{-1}$ and
$\rho_\infty = 10^{-21} \ {\rm{kg}}~{\rm{m}}^{-3}$. The polytropic
exponent, $\gamma$, has been set as $n \equiv (\gamma -1)^{-1}=1.61$. 
For these values of the physical constants, 
transonicity becomes apparent even at the very early stages of the 
evolution. This is shown in Fig.~\ref{f3} in which the velocity field 
(scaled as the Mach number) has been plotted after it has evolved for 
$4000$ seconds. The horizontal distance has been scaled by the sonic 
radius (which, going by equation~(\ref{rcrit}), effectively makes this 
scaling dependent on the fractal dimension), and it shows a saturation
behaviour for the inflow velocity at smaller length scales. This
saturation scale for the velocity field is roughly the same for
any value of $d$, although at larger length scales, a curve placed 
higher in the plot, corresponds to a higher value of $d$. Besides 
this, all curves, as it is apparent from the slope of each of them, 
display the same kind of power-law behaviour on larger length scales.  
All this is somewhat reminiscent of the growth processes exhibited 
by the ballistic deposition model, which is applied to generate 
a nonequilibrium interface~\citep{barstan}. 

The comparative properties of the velocity field, at smaller length
scales, for various values of $d$, have been exhibited in 
Fig.~\ref{f4}. Here the radial distance has been scaled in terms of 
the radius of the accretor, which in this case is $r_\odot$. What is 
interesting to note in this plot is that for all other conditions 
remaining the same, on length scales close to the accretor, solutions 
corresponding to lower values of the fractal dimension, $d$, grow 
faster in time than solutions with higher values of $d$. It is
possible to argue that this is exactly how it should be. 
The pressure of the infalling gas, in so far as it is connected to 
the density through a polytropic equation of state, builds up 
resistance against gravity, because of the growth of the density
field on small length scales. Transonicity can only be achieved when
gravity wins over pressure on length scales smaller than the sonic
radius. This will be all the more true near the accretor, where 
the velocity field will evolve under free-fall conditions, and, 
therefore, the more dilute the gas, the more efficient will be the 
drive towards the transonic state. Now a fractal medium may be 
viewed equivalently as a continuum with an effective lesser density. 
In this situation a system with 
a lesser value of $d$ will be more prone to losing against gravity
than a system with higher value of $d$, and so the race towards
transonicity will be more successful as $d$ decreases. It is 
exactly this state of affairs that Fig.~\ref{f4} graphically 
represents. 

\section{Concluding remarks}
\label{sec6}

An earlier work reported by~\citet{roy07} was carried out under the
implicit assumption that the accretion process would take place
transonically. The present treatment bears out this assumption
self-consistently. It was also discussed by~\citet{roy07} that 
the rate of accretion in a fractal medium can vary significantly 
from the~\citet{bon52} rate for more massive accretors. This is 
very much in conformity with the conclusions derived in this work,
through the numerical evolution of transonicity, and it would be
judicious to account for this fact, while studying the accretion 
of a fractal medium on to a black hole. 
Black hole accretion is necessarily transonic~\citep{skc90,skc96},
but even for accretion from a molecular cloud on to a star, in the
absence of any inner boundary condition being imposed on the flow,
the flow is expected to be transonic~\citep{pso80}. Therefore,  
whatever be the nature of the accretor, both transonicity and the 
quantitative modifications arising due to the fractal 
nature of the accreting medium, will be very much relevant for 
studies in spherically symmetric accretion. 

It has also been discussed here that transonic properties manifest
themselves more noticeably with an increase in the fractal 
properties of the flow. The dynamic evolution has shown that 
the growth rate of the velocity field (as scaled against the 
speed of acoustic propagation) becomes significantly higher in 
this case. This, of course, will have a direct bearing on the 
mass accretion rate, and it is worth conjecturing that there
might be some observational evidence for this kind of behaviour. 

\section*{Acknowledgements}

This research has made use of NASA's Astrophysics Data System. The
authors express gratitude to J. K. Bhattacharjee, J. N. 
Chengalur, T. Naskar and R. Nityananda for much encouragement and 
many helpful comments.

\bsp
\label{lastpage}
\end{document}